# Autonomous agile teams:
# Challenges and future directions for research

Viktoria Stray
University of Oslo, SINTEF
Norway
stray@ifi.uio.no

Nils Brede Moe
SINTEF
Norway
nilsm@sintef.no

Rashina Hoda
University of Auckland
New Zealand
r.hoda@auckland.ac.nz

## ABSTRACT
According to the principles articulated in the agile manifesto, motivated and empowered software developers—relying on technical excellence and simple designs—create business value by delivering working software to users at regular short intervals. These principles have spawned many practices. At the core of these practices is the idea of autonomous, self-managing, or self-organizing teams whose members work at a pace that sustains their creativity and productivity. This article summarizes the main challenges faced when implementing autonomous teams and the topics and research questions that future research should address.

## 1 INTRODUCTION

To succeed in complex environments, organizations developing software have to find ways to support and regulate their teams' autonomy according to the environmental demands and limitations. Furthermore, they have to take into consideration the degree of change and uncertainty, and that there is no one-size-fits-all autonomy approach [1]. We have researched the topic of autonomous teams in agile software development for some time [2-4], but the process of designing, supporting, and coaching autonomous agile teams is still not adequately addressed and understood in the context of software development organizations. Further, there is a need for new knowledge on how companies should organize for the right level of team autonomy and utilize autonomous agile teams to attain better performance, productivity, innovation, and value creation, and thus, increase competitiveness. One emerging question is "How can organizations give autonomous agile teams the authority and competence to set directions for new products so that organizations can deliver innovative and high-quality software more rapidly?"

To address the challenges of implementing autonomous teams, we organized the first international workshop on autonomous agile teams at XP 2018 (The 19th International Conference on Agile Software Development) to understand better the specific challenges and to suggest a future research agenda. The goal of the workshop was to facilitate knowledge sharing about the current practice of autonomous agile teams and deepen the knowledge about practices and strategies that enable autonomous teams. We use the label "autonomous teams" as a synonym for "self-organizing teams," "self-managing teams," and for "empowered teams."

### 1.1 Autonomous teams: the origins

The concept of autonomous teams is not new and has been studied and described from various perspectives in the past [2]. One of the earliest references dates back to the Tavistock group's study of English coal miners [5]. From this socio-technical perspective, autonomous teams were described as teams of 10-15 cross-trained individuals guided by the corporate vision, motivated by peer-pressure, and taking on the responsibilities of their former supervisors. From an organizational theory perspective, Morgan described four principles of self-organization [6] as (a) a *minimum critical specification* in which the senior management describes only the critical aspects required for teams to function effectively, (b) *requisite variety*, and (c) *redundancy of function*. These first three principles imply that teams should be internally composed of people with a variety of skills in order to cater to and effectively tackle the variety in their external environment and that individuals within the team should, to a high degree, be able to assist and replace each other as required. Finally, the fourth principle, (d) *learning to learn*, underpins the team's ability to engage in double-loop learning and drive continuous improvement. Another interesting legacy of autonomous teams can be traced back to the complex adaptive systems perspective, where Anderson and McMillan [7] define self-organizing teams as informal and temporary, not a part of the formal organization structure, formed spontaneously around issues, having a strong sense of shared purpose, deciding their own affairs, and including primary roles related to their tasks.

Perhaps the closest and most direct definition of autonomous teams as applied from outside software engineering into agile software development comes from the knowledge-management perspective. In their paper, "The New Product Development Game," Takeuchi and Nonaka [8] defined autonomous teams as those exhibiting three conditions, autonomy, cross-fertilization, and self-transcendence. *Autonomy* refers to the team being provided freedom by their management and being able and willing to assert that autonomy in their everyday work. *Cross-fertilization* refers to the team being formed from individuals with different specializations, behavior, and thinking so that regular interaction improves their understanding of each other's perspectives. *Self-transcendence* refers to regular self-evaluation and goal setting as well as designing better ways to achieve those goals. The paper by Takeuchi and Nonaka is known to have been the inspiration for the Scrum development method in which the self-organizing team is seen to be central to achieving agility.

Finally, the first introduction of autonomous teams into software engineering was made by way of the agile manifesto which cited self-organizing teams as the source of *"the best architectures, requirements, and designs"* [9]. Of the various agile methods, Scrum directly refers to and defines self-organizing teams, while eXtreme Programming refers to empowered teams. These definitions follow closely the attributes, conditions, and principles described from the knowledge management and organizational theory perspectives summarized above.

In addition to defining autonomous teams in terms of informal self-organizing roles [2] and using a teamwork model [10], research has attempted to describe a variety of challenges experienced by and arising from such teams. These include barriers to achieving cross-functionality at the organizational level [3] and project management challenges arising at the task, individual, team, and project levels due to increased involvement of teams in project management activities [11]. However, much remains to be explored in this context, at all levels.

## 1.2 Structure of the workshop

The workshop included two invited keynote presentations: "When is agile better? How the use of agile and autonomous teams affects success differently in different contexts" by Magne Jørgensen from Simula Research Laboratory, and "Organizing self-organization" by Rashina Hoda from the University of Auckland. Further, the workshop had four presentations by researchers who had had their papers peer-reviewed and accepted by members of the program committee. There were two highly interactive sessions where workshop techniques were used to generate discussions among all the participants. The workshop had group discussions on the main barriers to autonomous teams. Based on a grouping of these challenges, four topics emerged: 1) not having clear and common goals, 2) lack of trust, 3) too many dependencies to others, and 4) lack of coaching and organizational support. These barriers motivated for a discussion leading to a research agenda including the following topics: leadership, coordination, organizational context, team design, and team processes.

## 2 BARRIERS TO AUTONOMOUS AGILE TEAMS

Autonomous agile teams offer potential advantages over traditional managed software teams. However, team performance is complex, and an autonomous agile team's performance depends not only on the team's competence in managing and executing its work but also on the organizational context. Further, autonomy has a positive influence on team effectiveness when task interdependence is high and a negative effect when task interdependence is low [12]. Although most studies report positive effects from autonomous teams, some present a more mixed assessment; they can be difficult to implement and risk failure when used in inappropriate situations or without sufficient leadership and support. If the implementation of autonomy carries a cost greater than the benefit, or if the team cannot adequately take advantage of the autonomy, then the granting of autonomy is not only without benefit but could be harmful to team effectiveness [12]. The actual performance of an autonomous agile team depends not only on the competence of the team itself in managing and executing its work but also on the organizational context provided by management [13]. The following top barriers to autonomous teams were identified and then discussed during the workshop:

**Not having clear and common goals:** When there is ambiguity about the direction and what to achieve, people inside and outside the team spend time trying to figure out what is supposed to be accomplished, reducing the coordinated actions in the team.

**Lack of trust:** When there is a lack of trust within the team, team members do not commit to the team goals. When there is a lack of trust between the team and managers, managers demand more reporting and control while the team reduces their uptake of responsibility. External constraints such as fixed-price and fixed-scope contribute to this problem [14].

**Too many dependencies to others:** If the team needs to reach an agreement or synchronize deliverables with too many experts, managers, stakeholders, and other teams, their authority to make decisions regarding the development process, technology, architecture, and product is reduced. For example, the software architecture may limit team autonomy if the architecture results in many technical dependencies between teams, which requires a constant need for alignment and coordination [15].

**Lack of coaching and organizational support:** Autonomous teams are not created simply by exhorting democratic ideals, by tearing down organizational hierarchies, or by instituting one-person-one-vote decision-making processes. Further, teams often do not have the adequate resources and have difficulty finding a sustainable rhythm while avoiding excessive stress for the individuals. Managers can lack the training to coach for autonomy.

**Diversity in norms:** Norms are the informal rules that guide the team and regulate team members' behavior [16]. If norms are left to emerge on their own, they will often not support strategic thinking that is essential for autonomous teams.

The challenges identified above led to a discussion and a proposal for a research agenda described in the next section.

## 3 RESEARCH AGENDA

Five topics emerged at the workshop as important for future research to understand better how companies can effectively enable autonomous agile teams.

### 3.1 Leadership

Leaders have an important role in the life of autonomous teams. Leadership in autonomous teams is broadly distributed among a set of individuals instead of being centralized in the hands of a single individual acting in the role of a superior. For teams new to autonomy, leaders need to design and set the direction for the team, then help the team establish team norms and learn to learn, and finally be a coach for the team toward autonomy. Some traditional managers new to such a leadership role are unaware of the importance of such coaching and end up frequently asking for a detailed report of the team's progress, which ends up being detrimental to the team's autonomy [17]. Coaching autonomous teams in a large-scale setting is more complex than for single independent teams because of all the external dependencies and the need for external coordination. The following research questions were suggested:

- How to design, support, and coach autonomous teams?
- How do organizations build a capacity for shared leadership for autonomous teams in a multi-team setting?
- How can businesses (customers) and the team create a shared understanding of the business objectives, create a common "purpose"?

### 3.2 Coordination

Autonomous teams are severely challenged by the increasing need to coordinate their work and align their work processes with multiple experts, stakeholders, and other teams, for example, in a distributed or large-scale context. As the number of interdependences between people, tasks, their knowledge, technical systems, and other resources increases, the complexity of coordination increases exponentially in and outside of the autonomous agile teams. Common awareness or understanding of the current state of the team is essential to be able to coordinate effectively. Research has proposed different conceptual approaches, for example, complex adaptive systems (CAS) [7], transactive memory systems [18], and sensemaking [19]. The approaches contribute to the insight into how team members can coordinate their actions in response to what other team members and people outside the team are doing. The following research questions should be explored:

- How to coordinate dependencies among autonomous agile teams?
- How to create a common awareness and understanding of the current state of the team and its tasks?
- What are the effective intra- and inter-team coordination mechanisms for autonomous agile teams?
- How can dependencies between teams be reduced and managed?
- How can system architecture best support coordination of autonomous teams?

### 3.3 Organizational context

While many barriers for the effectiveness of autonomous agile teams lie at the team level and the leadership of the team, the organizational and environmental contexts also affect the success of such teams. Autonomous agile teams, especially those in large projects, face a variety of organizational constraints, for example, legislation, security, universal design, complex software architecture, legacy systems, and the need for standardization [20]. Further, the cultural and organizational contexts (including the organization's formal properties such as centralization, technology, and control systems) need more attention [21]. The following research questions were suggested.

- How to balance the need for alignment and team autonomy in complex organizations and multi-team environments?
- What is the right degree of autonomy in different contexts (and how to measure it)?
- How to enable organizations to adapt to autonomous teams?
- How to change the mindset of the wider organization to adopt agile autonomous teams?

### 3.4 Design of autonomous agile teams

Software companies face a growing environmental complexity that demands cross-functional autonomous teams. Often, there is a need to introduce DevOps, BizDev, or BizDevOps teams. The team's structure must support rather than impede the team, so there must be clear boundaries that distinguish members from non-members. Alignment between overall business strategies, innovation activities, development, and operations in autonomous agile teams is challenging. Putting all the needed skills within a team results in large teams which makes shared leadership and shared decision-making challenging. Furthermore, team members with different backgrounds often have different norms guiding them which may be a hindrance to being an effective agile team [24]. Future research should explore the following.

- What are the effective team structures for autonomous agile teams?
- What is the right team size for autonomous cross-functional teams?
- How should agile practices be adjusted to promote effectiveness in cross-functional teams?

### 3.5 Team processes

Autonomous teams stimulate participation and involvement, leading to team members developing an emotional attachment to the organization, greater commitment and motivation to perform, and a desire for responsibility [25]. Increased responsibility may lead to stress for the team members because they work at a high and self-transcending pace. However, a recent study found that self-organization showed a strong correlation to lower stress levels in agile teams and suggested that having a self-organizing team was the most important factor for lowering the level of stress [26].

Further, autonomous teams are prone to suffer from greater peer pressure than managed teams. At the same time, individuals need to be motivated and satisfied with their jobs by having control over their work and the scheduling and implementation of their tasks. There might be a conflict between the need for team and individual autonomy, especially in teams with a high degree of diversity. Newly formed autonomous teams frequently experience conflicts. Such teams often introduce roles such as team champion, tech liaison, and chief product owner [22] and mentor, coordinator, translator, promoter, and terminator [2]. New and unclear roles might result in misunderstandings and conflicts due not to interpersonal factors but because of team-related contextual factors such as unclear responsibilities. Therefore, the teams should have a formal structure for conflict resolution [23]. The following research questions were suggested.

- How to reduce stress in agile, highly motivated autonomous teams?
- How to resolve conflicts between roles and teams and between teams and managers?
- How to handle cultural differences and diversity in large-scale agile settings (e.g., age, gender, experience, culture, and field of expertise (biz vs dev))?
- What communication practices are best and when?
- How can communication tools such as Slack improve collaboration and coordination?

## 4  CONCLUSION

This paper gives an overview of what practitioners and researchers in the field of agile software development believe are emergent research themes for autonomous teams. Future research should explore the five identified topics in the research agenda: Leadership, coordination, organizational context, team design, and team processes. For the research agenda, we proposed several research questions to engage with to identify which factors increase, moderate, or limit the level of team autonomy and the effects of autonomy on team performance in agile software teams.

### ACKNOWLEDGEMENTS

The Research Council of Norway partially supported this work through grant 267704. Additional support was provided by the following companies: Kantega, Knowit, Storebrand, and Sbanken. We are very grateful to the following people who presented their papers at the workshop: Jan Henrik Gundelsby, Kjell Lundene, Lucas Gren, Per Lenberg, and Yngve Lindsjørn. Thanks to the program committee members for thorough reviews and all the workshop participants for engaging discussions. The participants were from a variety of institutions including: Accenture, Tampere University of Technology, Codecentric, Sbanken, Universidad Politécnica de Madrid, Agilcal AB, Bristol-Myers Squibb, Simula Research Laboratory, University of Auckland, Knowit, Johns Hopkins University, Flir, University of Oslo, Swift, Daimler, The Open University, Blekinge Technical University, SINTEF, Chalmers University of Technology, The University of Gothenburg, and The Norwegian Labor and Welfare Services.

# REFERENCES

[1] Chen, J., et al., *The relationship between team autonomy and new product development performance under different levels of technological turbulence.* Journal of Operations Management, 2015. **33–34**: p. 83-96.

[2] Hoda, R., J. Noble, and S. Marshall, *Self-organizing roles on agile software development teams.* IEEE Transactions on Software Engineering, 2013. **39**(3): p. 422-444.

[3] Moe, N.B., T. Dingsøyr, and T. Dybå, *Overcoming Barriers to Self-Management in Software Teams.* IEEE Software, 2009. **26**(6): p. 20-26.

[4] Stray, V.G., N.B. Moe, and T. Dingsoyr, *Challenges to Teamwork: A Multiple Case Study of Two Agile Teams*, in *Agile Processes in Software Engineering and Extreme Programming*, A. Sillitti, et al., Editors. 2011. p. 146-161.

[5] Trist, E., *The evolution of socio-technical systems: a conceptual framework and an action research program*, in *Occasional paper No 2*. 1981, Ontario Quality of Working Life Centre: Toronto, Ontario.

[6] Morgan, G., *Images of Organizations*. 2006, Thousand Oaks, CA: SAGE publications. 504.

[7] Anderson, C. and E. McMillan, *Of ants and men: Self-organized teams in human and insect organizations.* Emergence, 2003. **5**(2): p. 29-41.

[8] Takeuchi, H. and I. Nonaka, *The New New Product Development Game.* Harvard Business Review, 1986( 64): p. 137-146

[9] Agile Alliance. *Manifesto for Agile Software Development*. 2001 [cited 2010 August]; Available from: http://www.agilemanifesto.org/principles.html.

[10] Moe, N.B., T. Dingsøyr, and T. Dybå, *A teamwork model for understanding an agile team: A case study of a Scrum project.* Information and Software Technology, 2010. **52**(5): p. 480-491.

[11] Hoda, R. and L.K. Murugesan, *Multi-level agile project management challenges: A self-organizing team perspective.* Journal of Systems and Software, 2016. **117**: p. 245-257.

[12] Langfred, C.W., *Work-Group Design and Autonomy:A Field Study of the Interaction Between Task Interdependence and Group Autonomy.* Small Group Research, 2000. **31**(1): p. 54-70.

[13] Hoda, R. and J. Noble. *Becoming Agile: A Grounded Theory of Agile Transitions in Practice*. in *2017 IEEE/ACM 39th International Conference on Software Engineering (ICSE)*. 2017.

[14] Lindsjørn, Y. and R. Moustafa. *Challenges with lack of trust in agile projects with autonomous teams and fixed-priced contracts*. in *Proceedings of the Scientific Workshops of XP2018*. 2018. Porto, Portugal: ACM.

[15] Gundelsby, J.H. *Enabling autonomous teams in large-scale agile through architectural principles*. in *Proceedings of the Scientific Workshops of XP2018*. 2018. Porto, Portugal: ACM.

[16] Forsyth, D.R., *Group dynamics*. 2018: Cengage Learning.

[17] Stray, V., N. B. Moe and D.I.K. Sjøberg, *Daily Stand-Up Meetings: Start Breaking the Rules.* IEEE Software, 2018 (in press).

[18] Wegner, D.M., *Transactive memory: A contemporary analysis of the group mind*, in *Theories of group behavior*. 1987, Springer. p. 185-208.

[19] Weick, K.E., *Sensemaking in organizations*. Vol. 3. 1995: Sage.

[20] Lundene, K. and P. Mohagheghi. *How Autonomy Emerges as Agile Cross-Functional Teams Mature*. in *Proceedings of the Scientific Workshops of XP2018*. 2018. Porto, Portugal: ACM.

[21] Oldham, G.R. and J.R. Hackman, *Not what it was and not what it will be: The future of job design research.* Journal of organizational behavior, 2010. **31**(2-3): p. 463-479.

[22] Nyrud, H. and V. Stray. *Inter-Team Coordination Mechanisms in Large-Scale Agile*. in *Proceedings of the Scientific Workshop Proceedings of XP2017*. 2017. ACM.

[23] Gren, L. and P. Lenberg. *The Importance of Conflict Resolution Techniques in Autonomous Agile Teams*. in *Proceedings of the Scientific Workshops of XP2018*. 2018. Porto, Portugal: ACM.

[24] Stray, V., T.E. Fægri, and N.B. Moe. *Exploring norms in agile software teams*. in *International Conference on Product-Focused Software Process Improvement*. 2016. Springer.

[25] Fenton-O'Creevy, M., *Employee involvement and the middle manager: evidence from a survey of organizations.* Journal of Organizational Behavior, 1998: p. 67-84.

[26] Meier, A., et al. *Stress in Agile Software Development: Practices and Outcomes*. 2018. Cham: Springer International Publishing.